\documentclass[12pt,preprint]{aastex}

\begin{document}

\title{Exponential Disks from Stellar Scattering: III. Stochastic Models}

\author{Bruce G. Elmegreen\altaffilmark{1}, Curtis Struck\altaffilmark{2}}

\altaffiltext{1}{IBM Research Division, T.J. Watson Research Center, P.O. Box 218,
Yorktown Heights, NY 10598; bge@watson.ibm.com}

\altaffiltext{2}{Department of Physics \& Astronomy, Iowa State University, Ames, IA
50011; struck@iastate.edu}

\begin{abstract}
Stellar scattering off irregularities in a galaxy disk has been shown to make an
exponential radial profile, but no fundamental reason for this has been suggested. Here
we show that exponentials are mathematically expected from random scattering in a disk
when there is a slight inward bias in the scattering probability.  Such a bias was
present in our previous scattering experiments that formed exponential profiles. Double
exponentials can arise when the bias varies with radius. This is a fundamental property
of scattering and may explain why piece-wise exponential profiles are ubiquitous in
galaxies, even after minor mergers and other disruptive events.
\end{abstract}


\section{Introduction}
\label{intro}

The exponential shape for the radial profiles of galaxy disks \citep{dv59} has never been
explained at a fundamental level. The initial mass and angular momentum distribution in
the gaseous halo plays a role early-on \citep{eggen62,mestel63,fall80}, as they lead to a
nearly exponential shape during collapse if angular momentum is conserved
\citep{freeman70}. Numerical simulations confirm this result even with some
redistribution of angular momentum \citep[e.g.,][]{dalcanton97,governato07,
foyle08,sanchez09, cooper13, aumer13, aumer13b, stinson13,martig14,herpich15,minchev15,
rathaus16}. Initial conditions also seem involved for the far-outer gas disks of HI-rich
galaxies, which have exponential profiles with a universal scale length when normalized
to the radius where the HI surface density is $1\;M_\odot$ pc$^{-2}$ \citep{wang14}.
\cite{bigiel12} found a universal exponential gas profile when normalized to $R_{25}$,
the radius at 25 magnitudes per square arcsec in the V band.

Disk stars also move radially after they form because of bulk motions in the presence of
torques \citep{hohl71}, guiding center migrations that preserve nearly circular orbits
\citep{se02,roskar08,veraciro14}, and strong scatterings that make eccentric orbits
\citep{bournaud07}. These motions can produce exponential profiles too. Numerical
simulations show piece-wise exponentials after galaxy mergers
\citep{younger07,borlaff14,athan16}, and galactic bars \citep{debattista06} or bars
coupled with spirals \citep{minchev12} produce exponentials too. The bar and spiral
profiles are consistent with observations of the locations of break radii between
exponential segments, which tend to be at the outer Lindblad resonances of these patterns
\citep{munozmateos13,laine14}. When there are no obvious perturbations, scattering off
interstellar clouds and holes can still make exponentials \citep{elmegreen13,struck16},
which are observed in dwarf Irregular galaxies too \citep{herrmann13}.

The underlying reason for exponential shapes in stellar scattering models has not been
adequately investigated. This result was robust in our earlier papers, with exponential
radial profiles appearing in two and three-dimensional models from initially uniform
stellar distributions with randomly positioned scattering sites. The stellar particles
were not self-gravitating so that spiral arms would be avoided, and a fixed potential for
the orbits was assumed to avoid stellar momentum exchange with anything other than the
scattering sites. These are the most basic ingredients for any disk model, so whatever
drives evolution in them should also be present to some degree in more complete models.

Here we consider an interesting aspect of random scattering in a disk that may be
relevant to the galaxy problem. This involves stochastic scattering with a slight inward
bias, which always makes a radial profile that is essentially the same as an exponential.
We also give evidence from our published scattering models that even though there was an
overall expansion of the disk as the exponential grew in length, there was still a slight
inward bias for each scattering event. Whether this model applies to real galaxies is not
known, so we offer some observational tests for future research.

In what follows, Section \ref{line} shows the change from a Gaussian to an exponential
profile for one-dimensional scattering on a line when there is a biased probability
toward the direction of a reflecting wall. Section \ref{disk} then considers
two-dimensional scattering in a disk, which requires only the inward bias without the
reflecting wall to make an exponential. Analytical models for these scattering profiles
are in Section \ref{analytic}. Double exponentials as in Type II and III galaxies are
discussed in Section \ref{double}. Section \ref{models} shows the scattering bias in our
previous 3D model for comparison with the stochastic theory. Section \ref{tests} offers
some tests on the model. A summary is in Section \ref{sum}.

\section{Scattering along a Line}
\label{line}

\begin{figure}
\epsscale{1}
\plotone{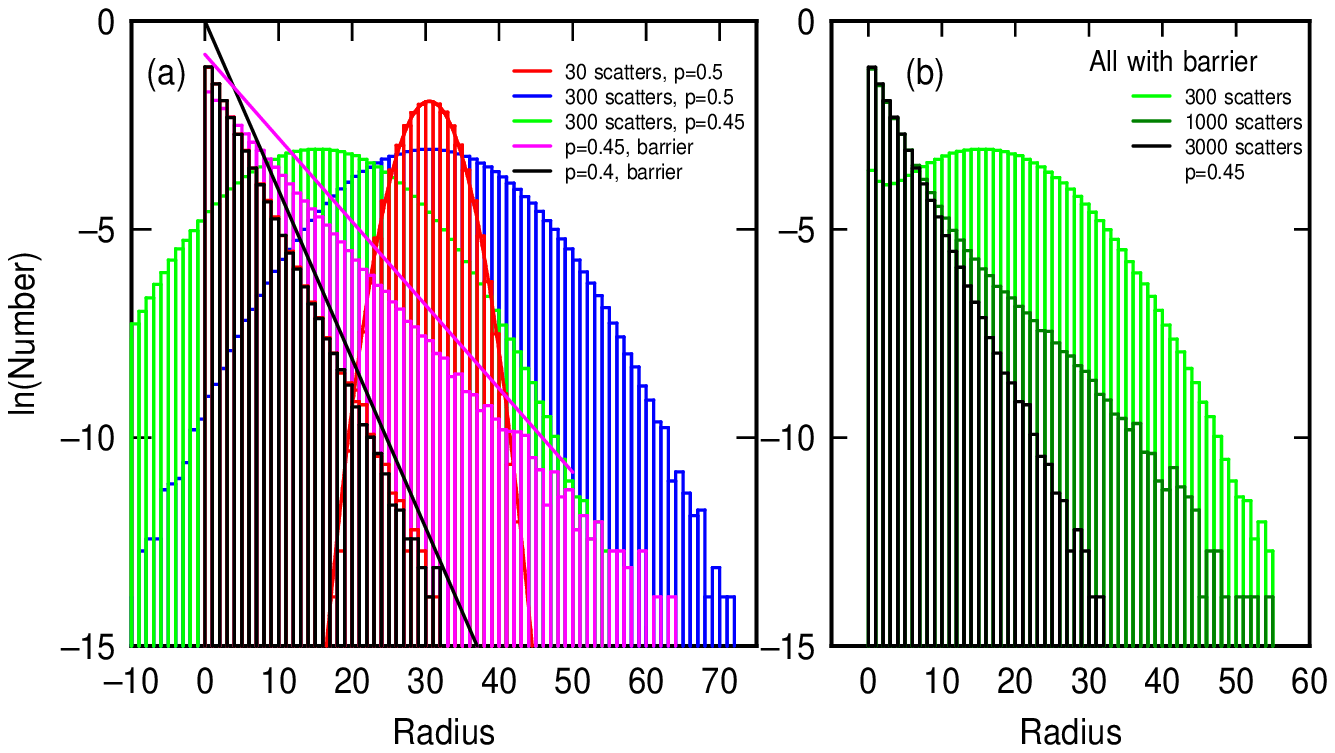}
\caption{(a) One-dimensional distribution of $10^6$ particles launched from position 30
and scattered $N=30$ times (red) and 300 times (blue) with equal probabilities of scattering
to the left ($q=0.5$) and right ($p=0.5$). The red histogram has a superposed Gaussian
with the theoretical dispersion $(Npq)^{0.5}$. The distribution shifts to the left (green)
when there is a bias in scattering that direction, {using $p=0.45$ and $q=0.55$}. The magenta and black
distributions also launch particles from position 30 but they have a reflecting barrier
at position 0, which turns the Gaussian shape into an exponential. There are $N=3000$
scatterings per particle in these two cases.  The step size for the scattering without a reflecting
barrier is 1/2 unit, and the step size for scattering with a reflecting barrier is 1 unit.
The scale length of the
exponential is $1/\ln(q/p)$ in this case (Eq. 13),
as indicated by the lines running offset and parallel to the
histograms. (b) Steps in the development of an exponential for $p=0.45$
as the number of scatterings per particle increases from 300 to 1000 to 3000.
The step size is 1/2 unit, as for the non-reflective cases in (a).
\label{pinball4two}}
\end{figure}

A convenient model for demonstrating the effects of scattering is a random walk along a
line with steps to the left or right having certain fixed probabilities. Figure
\ref{pinball4two}a shows a normalized histogram in red with the usual Gaussian shape made
by scattering particles away from a starting point at $x=30$ with an equal probability to
the left ($q=0.5$) and right ($p=0.5$). The jumping distance per scatter is 0.5 units
(see Sect. \ref{analytic}). The theoretically expected dispersion for this classical
problem is $(Npq)^{0.5}=2.73$ for $N=30$ scatterings per point in this case. A
theoretical Gaussian with this dispersion is shown superposed on the red histogram with a
perfect fit. The number of particles used is $10^6$, and the area is normalized to unity.

The blue histogram in Figure \ref{pinball4two}a shows another case with particles
launched from $x=30$ and $p=0.5$, but now there are 300 scatterings per point, giving a
larger dispersion.  The green histogram is the same as the blue one but it has a bias to
the left with $q=0.55>p=0.45$. The bias moves the Gaussian to the left and shows no
indication of the launching point, which is still at $x=30$. Note that the Gaussian
histograms cover both positive and negative $x$.

The shape of the histogram in Figure \ref{pinball4two}a changes significantly when there
is a reflecting barrier in the direction of the bias. The magenta histogram shows a case
like the green one, with $10^6$ particles launched from $x=30$ and a bias $p=0.45$, but
now there is a reflecting barrier at $x=0$. That is, any particle that reaches $x=1$ and
is randomly directed toward the left is automatically returned to $x=1$.  The result of
this reflection is to convert a Gaussian into an exponential. The scale length of the
exponential is $1/\ln(q/p)=4.98$, as derived in Section \ref{analytic}; a theoretical
exponential with this scale length is shown by the straight magenta line, which runs
parallel to the histogram, as expected. The jump distance in this case is 1 unit (see
Sect. \ref{analytic}). The black histogram is made the same way as the magenta one, but
with $p=0.4$, which has a theoretical scale length of $2.466$, as shown by the black
line. Unlike the case for the Gaussians with no barrier, the exponentials do not change
their shape as the number of scatterings per particle increases, once this number reaches
a high enough value. Figure \ref{pinball4two}a has $N=3000$ scatterings per particle for
the two exponential cases. Increasing the number of particles affects all of the
normalized histograms in the same way, by extending their reach to lower values on the
ordinate (i.e., they go down to a value of $1/N$, at which point only one particle is
present in the histogram bin.) Note that the launching point of $x=30$ does not show up
in the exponential profiles because they have reached an equilibrium shape at high $N$.

Figure \ref{pinball4two}b illustrates the development of the exponential for $p=0.45$ as
the number of scatterings per particle increases, all with $10^6$ particles again. The
step size is 0.5, as in the Gaussians for figure \ref{pinball4two}a. For small numbers of
scatterings per particle, the reflecting barrier is rarely sampled and a Gaussian
histogram like the green one in Figure \ref{pinball4two}a appears. As each particle
scatters more, the counts on the left climb up the barrier and the exponential appears.
An explanation for this is in Section \ref{analytic}.

The black histogram with $p=0.4$ in Figure \ref{pinball4two}a looks the same as the black
histogram with $p=0.45$ in Figure \ref{pinball4two}b although the two curves have
different biases. This similarity is coincidental. The exponential cases in Figure
\ref{pinball4two}a have unit steps for jumps in position, making equation
(\ref{eq:solution}) below relevant, while all cases in Figure \ref{pinball4two}b have
half steps for the jumps in position to make a fair comparison with the green Gaussian in
Figure \ref{pinball4two}a, which also has half-step jumps. When the step size is 0.5
unit, the scale length is half as much, as noted in the text below equation
(\ref{eq:solution}). Thus the scale length for the black histogram in Figure
\ref{pinball4two}a is $1/\ln(0.6/0.4)=2.466$ as mentioned above, and the scale length for
the black histogram in Figure \ref{pinball4two}b is $0.5/\ln(0.55/0.45)=2.492$, which is
about the same.

\section{Scattering in a Disk}
\label{disk}

\subsection{Method} \label{method}

\begin{figure}
\epsscale{1.}
\plotone{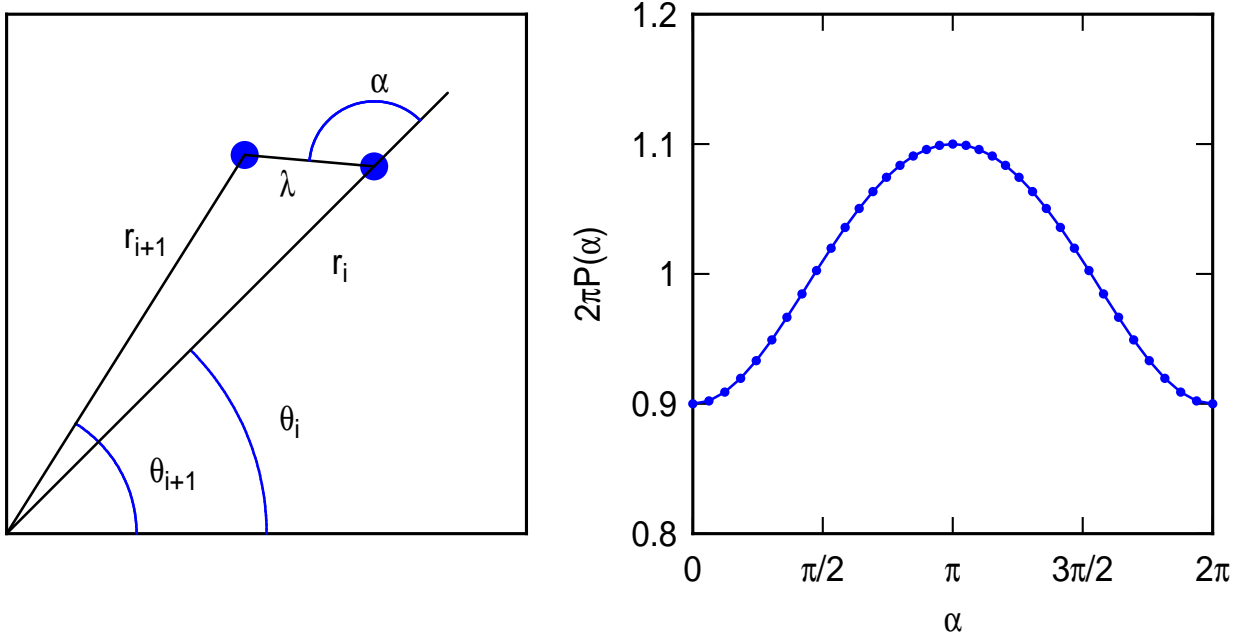}
\caption{(left) Nomenclature for radii and angles in the two-dimensional scattering
experiments. The particle scatters from position $i$ to $i+1$. (right) Distribution
function of the scattering angle $\alpha$ showing a bias with $b=0.1$ toward the center of the
coordinate system ($\alpha=\pi$).  \label{pinball_makefigure}}
\end{figure}

Scattering in a disk geometry involves the radial position of a particle and two angles,
one for the particle position in the disk and another for the direction of scattering. We
consider a circular coordinate system $(r,\theta)$ and some current position of a
particle, $(r_i,\theta_i)$ (see Fig. \ref{pinball_makefigure}). The particle is scattered
for a unit distance $\lambda=1$ in some direction $\alpha$ measured from the direction
toward larger radii; $\alpha$ increases as the angle around the particle increases
counter-clockwise.

This angle $\alpha$ is chosen with a bias toward the center of the coordinate system by
picking a random number $\zeta$ between 0 and 1 and letting a generating angle
$\alpha^\prime$ be equal to $2\pi \zeta$, which means that $\alpha^\prime$ is distributed
uniformly. The actual scattering angle $\alpha$ is not distributed uniformly, but has an
inward bias that comes from the generating angle $\alpha^\prime$. To generate this bias,
we take
\begin{equation}
y=\sin(\alpha^\prime)\;;\;x=\cos(\alpha^\prime)-b
\label{biasb}
\end{equation}
for $b>0$ a bias amount. The inwardly biased angle for next scattering is then
\begin{equation}
\alpha={\rm arctan}(y/x).
\end{equation}
The next radial position is determined from the triangle made by the current radial
position and the scattering jump, using the trigonometric theory of cosines:
\begin{equation}
r_{i+1}=\sqrt{r_i^2+\lambda^2+2\lambda r_i \cos(\alpha)}.
\label{alphaeq}
\end{equation}
The next angular position in the coordinate system, $\theta_{i+1}$, follows by first
introducing the angle $p$ from the trigonometric theory of sines, where
\begin{equation} \sin(p)=\lambda \sin(\alpha)/r_{i+1}
\end{equation}
and then
\begin{equation}
\theta_{i+1}=\theta_i+p.
\label{eq:theta}
\end{equation}

The right-hand side of Figure \ref{pinball_makefigure} shows the distribution of
scattering probabilities, $P(\alpha)$, as a function of the generating angle
$\alpha^\prime$ around the particle with $b=0.1$. It is derived by setting
$P(\alpha)d\alpha=P^\prime(\alpha^\prime)d\alpha^\prime$ with a constant distribution
function for $\alpha^\prime$, namely, $P^\prime(\alpha^\prime)=1/2\pi$. This gives
$P(\alpha)$ in terms of $\alpha^\prime$ from the derivative,
\begin{equation}
P(\alpha)={1\over{2\pi}} {{d\alpha^\prime}\over{d\alpha}}={{1}\over{2\pi}}
{{1-2b\cos\alpha^\prime+b^2}\over{1-b\cos\alpha^\prime}},
\label{probability}
\end{equation}
and then we convert $\alpha^\prime$ into $\alpha$ for the value on the abscissa, as
above.  The dots in the plot are the equal intervals of $\alpha^\prime$ used to evaluate
$P(\alpha)$ for the figure. The probability is higher for $\alpha\sim\pi$, which
corresponds to an inward direction.

\subsection{Exponentials from an inward scattering bias}
\label{subdisk}

\begin{figure}
\epsscale{1.}
\plotone{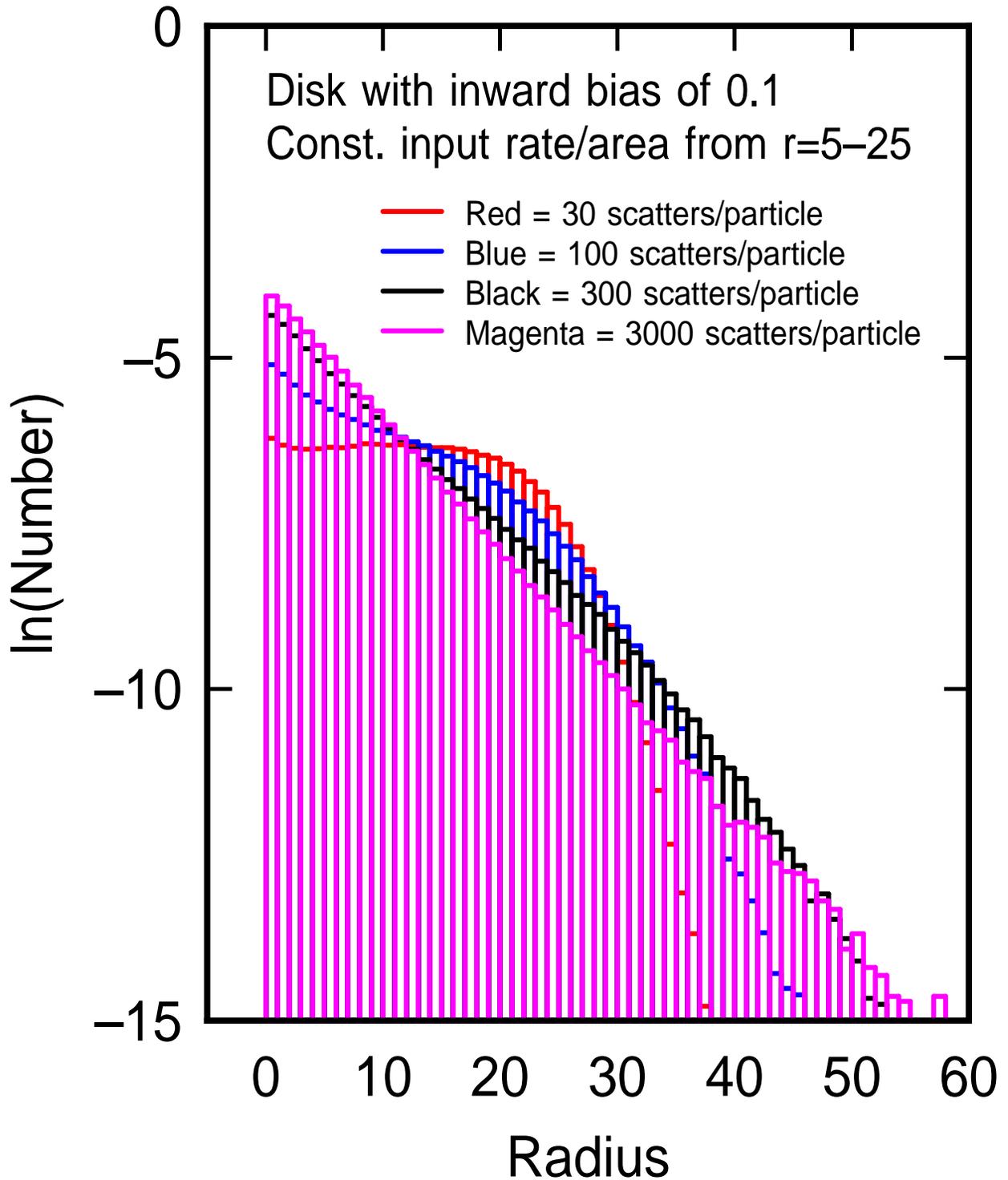}
\caption{Scattering in a disk with increasing number of scatterings per particle
in the 4 histograms showing convergence to an exponential radial profile.
Particles are launched with a uniform rate per unit area
from a radius of 5 to a radius of 25. A slight inward bias is imposed on the
scattering angle with the parameter $b=0.1$.   \label{pinball6}}
\end{figure}

Figure \ref{pinball6} shows histograms of the radial positions of 315000 particles that
follow from scattering in 2 dimensions with a bias ($b=0.1$, see eq. (\ref{biasb}))
directed toward the coordinate center $r=0$.  A histogram of azimuthal positions,
$\theta$, from equation (\ref{eq:theta}) is flat because the particles scatter all over
the disk (not shown). The launching radius ranges from 5 to 25 distance units, with the
number of particles launched increasing linearly with distance, to mimic a constant
surface density in this radial range.  The resulting distribution is again exponential
for a large number of scatterings per point because the origin of the coordinate system
acts like a reflecting barrier. This is because points near the center that scatter
toward smaller radius end up on the other side of the center at positive radius again.
There is no imposed actual reflection, but just a natural impossibility of going to a
negative radius.  The four histograms in Figure \ref{pinball6} have different numbers of
scatterings per particle. As the number increases, the exponential appears. The steepest
exponential in the figure is essentially the equilibrium distribution; decreasing the
number of scatterings from 3000 to 1000 (not shown) increases the scale length by only
3.8\%.

\begin{figure}
\epsscale{1.}
\plotone{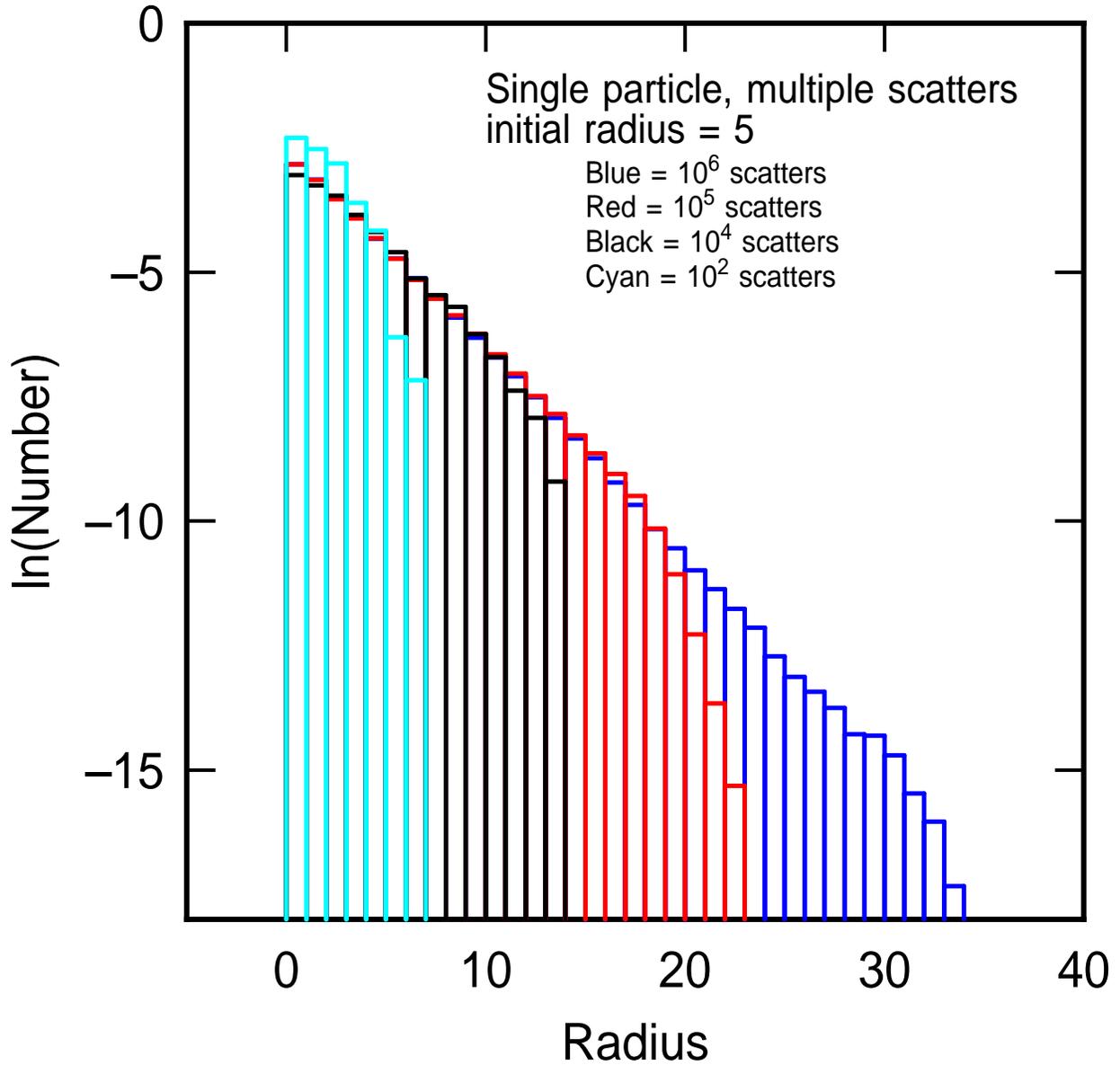}
\caption{Four distribution functions show the positions occupied by a single
particle that scatters a various number of times from $10^2$ (cyan) to $10^6$ (blue).
Each case starts with a particle at a radius of 5 and has an inward bias given by
$b=0.2$. As the number
of scatterings increases, the distribution function of radii covered by
that particle broadens but maintains a fixed exponential slope for small radii,
which are well sampled. \label{pinball7b}}
\end{figure}

Figure \ref{pinball7b} shows normalized histograms of radial positions of a single
particle that scatters various numbers of times from $N=10^3$ to $N=10^6$, always
starting at a radial position $x=5$ and with an inward bias of $b=0.2$. As the number of
scatterings increases, the limits to the exponential become broader, but there is always
about the same slope for small radii, converging to the theoretically predicted value for
the assumed bias when $N$ is sufficiently large. Note that the radial spread in the
histograms is not from an outward bias, but simply from scattering multiple times.

\section{Analytic Theory}
\label{analytic}
\subsection{Scattering on a line with no reflecting barrier}
The widths and slopes of the equilibrium distributions of particle positions are derived
here. Consider first a discrete model of scattering by fixed increments to the left or
right on a line. The probability of scattering to the left is $q>0.5$ and the probability
of scattering to the right is $p<0.5$, where $p+q=1$. The distance of scattering is $1/2$
so the difference between the two possible scattered positions is 1, i.e., one cell width
in a histogram. The cells alternate alignment along the half step and the full step. This
is the standard model for a Galton box or bean machine \citep{trow}. We imagine that a
new particle is introduced at position $i=0$ at each time step, and that all existing
particles at their integer positions $i$ scatter to the left or right with these
probabilities.

Without a reflecting barrier, the number of particles at position i after N steps is, on
average, equal to the number in the previous step, $N-1$, at position $i-1/2$ times the
probability that these particles scattered 1/2 step to the right, which is $p$, plus the
number in the previous step at $i+1/2$ which scattered 1/2 step to the left, which is
$q$:
\begin{equation}
\Phi_N(i)=\Phi_{N-1}(i-1/2)p+\Phi_{N-1}(i+1/2)q.
\label{gaussscat}
\end{equation}
In this case, $i$ can be either positive or negative. Iterating this back to $N=0$ on the
right gives the usual result from the coefficients of the binomial expansion of
$(p+q)^i$, which in the limit of large $N$ and normalized to unit area is a Gaussian
\begin{equation}
\Phi_N(i)={1\over{\sqrt{2\pi} Npq}}\exp\left(-0.5(i/Npq)^2\right)
\end{equation}
This explains the Gaussian distributions obtained above in the case of no reflecting
barrier. The center of the Gaussian occurs at the cell position where the particles
enter.

\subsection{Scattering on a line with a reflecting barrier}
With a reflecting barrier at $x=0$ and with $q>p$, the probability distribution changes
into an exponential. This may be seen in the same way by considering a particle
introduced at any position $i$ at each time step and following the left or right
scatterings of all existing particles, as above. The difference is that now the number at
position $i=0$ is the sum of the number in the previous step at $i=0$ times the
probability of a leftward motion, in which case the particle does not move at all because
of the barrier, plus the number at $i=1$ times the probability of a leftward motion,
\begin{equation}
\Phi_N(0)=\Phi_{N-1}(0)q+\Phi_{N-1}(1)q
\label{expscat}
\end{equation}
All other probabilities, $\Phi_{N}(i)$, are as in equation (\ref{gaussscat}) because they
do not feel the barrier. A minor detail is that we take aligned cells now for each step
so there is always a reflecting barrier at the lower end ($i=0$), and we therefore take a
unit length jump to the left and right. Thus we write in analogy to equation
(\ref{gaussscat}),
\begin{equation}
\Phi_N(i)=\Phi_{N-1}(i-1)p+\Phi_{N-1}(i+1)q.
\label{gaussscat2}
\end{equation}
Iterating Equation \ref{expscat} back to the first step now gives a leading term
$\Phi_N(0)=q^N$ and other terms multiplied by some combination of $q^kp^{N-k}$ for $k=1$
to $N-1$. This is because all counts $\Phi_{N}(i>0)$ have to contain at least one product
with $p$ in order to get the particles somewhere to the right of $i=0$. Similarly, one
can derive the last term $\Phi_N(N)=p^N$. Experimentation with real examples readily
shows that the distribution function $\Phi_N(i)$ has approximately equal factors ($<1$)
incrementing towards higher $i$, which means it is an exponential function. Taking just
the endpoints, we derive the scale length as follows:
\begin{equation}
\ln\Phi_{N}(N)-\ln\Phi_{N}(0)=N\ln(p)-N\ln(q)=-N\ln(q/p)
\end{equation}
or
\begin{equation}
\Delta\ln\Phi_N/\Delta i=-\ln(q/p)
\end{equation}
which has the solution in the continuum limit for $0\leq i\leq N$
\begin{equation} \Phi_N(i)=\exp\left(-i\ln(q/p)\right).\label{eq:solution}\end{equation}
Thus the exponential scale length is $1/\ln(q/p)$. Figure \ref{pinball4two}a has lines
with slope of $-\ln(q/p)$, showing agreement between the stochastically determined
histogram and this analytical result. Recall that without the barrier in the first
example, we took alternately aligned cells with a half length jump.  If we took a half
length jump in the exponential case, the scale length would be half as large,
$0.5/\ln(q/p)$. This is the case in Figure \ref{pinball4two}b to facilitate comparison
with the non-reflecting distribution functions, i.e., the Gaussians, in Figure
\ref{pinball4two}a.

\subsection{Scattering in a disk}
\label{analysisdisk}

In a circular geometry, the number of particles in radial bins may be written as in
equation (\ref{gaussscat}), but the surface density is more important, and this involves
a radial coordinate to keep track of the area in each radial interval. We introduce the
surface density $\Sigma$ and write the total number $\Phi$ as above with $\Phi\propto
r\Sigma$.  Knowing that the exponential reaches an equilibrium profile independent of the
number of scatterings, $N$ for large $N$ (Sect. \ref{line}), we can write equation
(\ref{gaussscat}) as an equilibrium equation for the number of particles, considering an
average step size $\lambda$ for each scattering,
\begin{equation}
\Phi(r-\lambda/2)p=\Phi(r+\lambda/2)q
\label{simpleform}
\end{equation}
This equation states that at some position $r$, the number of outward jumps at
probability $p$ from a slightly smaller radius $r-\lambda/2$ equals the number of inward
jumps at probability $q$ from a slightly larger radius $r+\lambda/2$, per unit time. We
let the excess probability of inward to outward scatterings be $\epsilon=q-p$. Setting
$\Phi(r-\lambda/2)=\Phi(r)-0.5\lambda d\Phi(r)/dr$ and the same for $\Phi(r+\lambda/2)$,
equation (\ref{simpleform}) becomes
\begin{equation}
{{d\Phi(r)}\over{dr}}=-{{2\epsilon}\over{\lambda}}\Phi(r),
\label{diffeq}
\end{equation}
which has the solution
\begin{equation}
\Phi(r)=\Phi_0\exp(-2\epsilon r/\lambda).
\label{solution}
\end{equation}
This is an exponential with a scale length of $r_{\rm D}=0.5\lambda/\epsilon$.

We get the same result if we consider the scattering more carefully in two dimensions,
using the angle $\alpha$ introduced above in equation (\ref{alphaeq}). Then the balance
between outward and inward scattering at position $r$ is written,
\begin{equation}
{1\over\pi} \int_{-\pi/2}^{\pi/2} \Phi(r-0.5\lambda\cos\alpha)(1-\epsilon\cos\alpha)d\alpha
= {1\over\pi} \int_{-\pi/2}^{\pi/2} \Phi(r+0.5\lambda\cos\alpha) (1+\epsilon\cos\alpha)d\alpha
\end{equation}
which reduces to equation (\ref{diffeq}) and solution (\ref{solution}). This integral
form considers all intermediate scattering angles. Note that the angle-dependent
scattering probability given by equation (\ref{biasb}) may be written approximately as
$(1-b\cos\alpha)/2\pi$, so $\epsilon\approx b$ in the present notation. Both of these
expressions are not meant to reflect the actual scattering distribution function for real
stars in a galaxy, but are the lowest order approximation to that function which
preserves a directional bias.

The surface density $\Sigma$ defined above is proportional to $\Phi(r)/r$, which varies
as $e^{-r/r_{\rm D}}/r$ using $r_{\rm D}=0.5\lambda/\epsilon$. For $r>r_{\rm D}$, this
function becomes indistinguishable from an exponential with the same scale length. The
inverse scale length in equation (\ref{solution}), $2\epsilon$, agrees fairly well with
the slope of the histogram in Figure \ref{pinball6}, considering that $\lambda=1$ in our
discrete model. This slope is $-0.187$ outside $r=10$ for inward bias $b=\epsilon=0.1$
and $N=3000$ scatterings per particle. For $b=0.05$ (not shown in the figure) the slope
is $-0.113$, and for $b=0.2$, the slope is $-0.388$, both with $N=3000$.

\section{Double exponentials}
\label{double}

Galaxy disks often show double exponential profiles with a downward bend at mid-radius
for Type II and an upward band for Type III \citep[e.g.,][]{pohlen06}. For Milky Way size
disks, this bend could be the result of a change in stellar populations with old stars in
the outer parts, as determined by color gradients \citep{bakos08,yoachim12} or age
gradients \citep{roediger12,ruiz16}. Age gradients are also present even in Type I disks,
which suggests that stellar scattering is present in all disks.  Because of these age
gradients, the mass profiles can be a single exponential even if the light profile is
Type II \citep{zheng15}. The average of a large number of light profiles at $3.6\;\mu$,
which is close to the mass profile at this wavelength, is a slightly down-bending double
exponential \citep{munozmateos13b}.

Although there are reasonable dynamical models that can produce a double profile in the
mass, purely stochastic scattering can do this too, making exponentials for each one, if
there is a difference in the scattering bias for the inner and outer regions. If the
outer part of a disk has a larger inward bias than the inner part, then the slope in the
outer part will be steeper. A physical reason for this might be that the outer parts have
only gas irregularities which scatter with an inward bias, while the inner parts have
both gas irregularities with an inward bias plus spirals or bars with an outward bias
from torques. Similarly, if the outer part of the disk has a smaller inward bias than the
inner parts, perhaps because of weaker gas clumps in the outer parts, then the purely
stochastic slope in the outer part will be shallower. In both cases, each segment will be
approximately exponential because of the scattering process.

\begin{figure}
\epsscale{1.}
\plotone{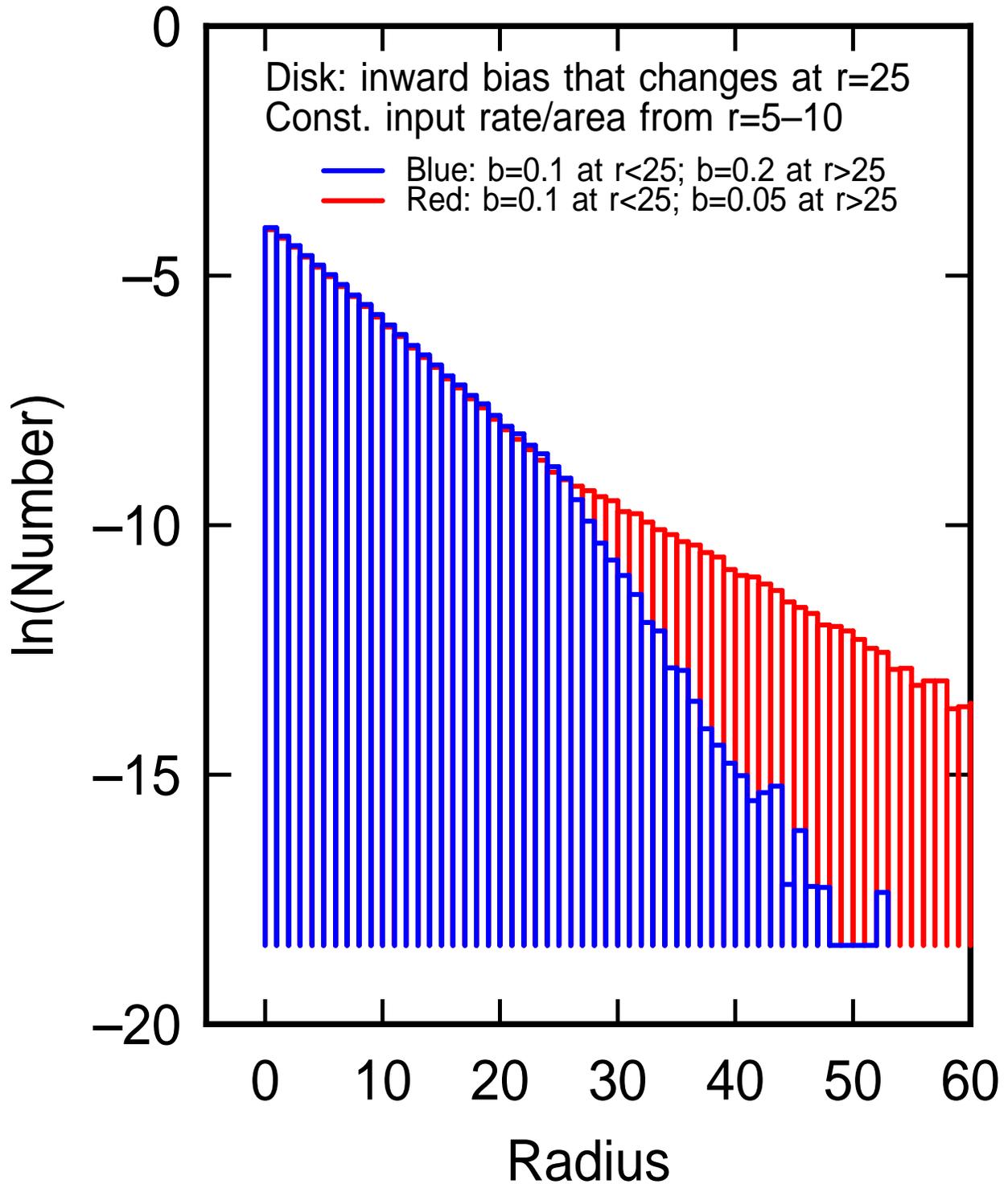}
\caption{Double exponentials made from scattering 330,000 particles launched with a
uniform surface density between radii of 5 and 10, and with
1000 scatterings per particle. Both histograms have an inward bias of $b=0.1$
for radii less than 25. The red and blue histograms have smaller, $b=0.05$, and larger,
$b=0.2$, inward biases
outside that radius, producing double exponential forms of Types III and II, respectively.
\label{pinball6b}}
\end{figure}

Figure \ref{pinball6b} shows the results of scattering experiments in these two cases.
Both histograms use 330000 particles with 1000 scatterings each. The particles are
launched between radii of $r=5$ and 10, with a number increasing linearly with radius to
mimic a constant surface density. Both histograms also have a scattering bias from
equation (\ref{biasb}), $b=0.1$, inside a radius of $r=25$, but the blue histogram has
$b=0.2$ outside $r=25$ while the red histogram has $b=0.05$ outside $r=25$. Each
exponential is formed separately and locally, although the particles scatter over the
disk. A gradually changing bias will produce a profile with more continuous curvature.

\section{Numerical experiments}
\label{models}

\cite{elmegreen13} and \cite{struck16} ran simulations with non-self-gravitating test
particles orbiting in a dwarf-irregular type galaxy potential with the addition of
gravitational scattering centers from mass irregularities that are analogous to large
clouds and holes in the interstellar medium. The test particles moved around in the disk
because of the scattering and formed a stable exponential radial profile from an
initially flat profile. We suspect that the exponentials in those papers have the same
cause as the stochastic exponentials studied here, so we measured the scattering bias in
the three-dimensional model of the second paper to see if it is directed inward.

\begin{figure}
\epsscale{1.4}
\plotone{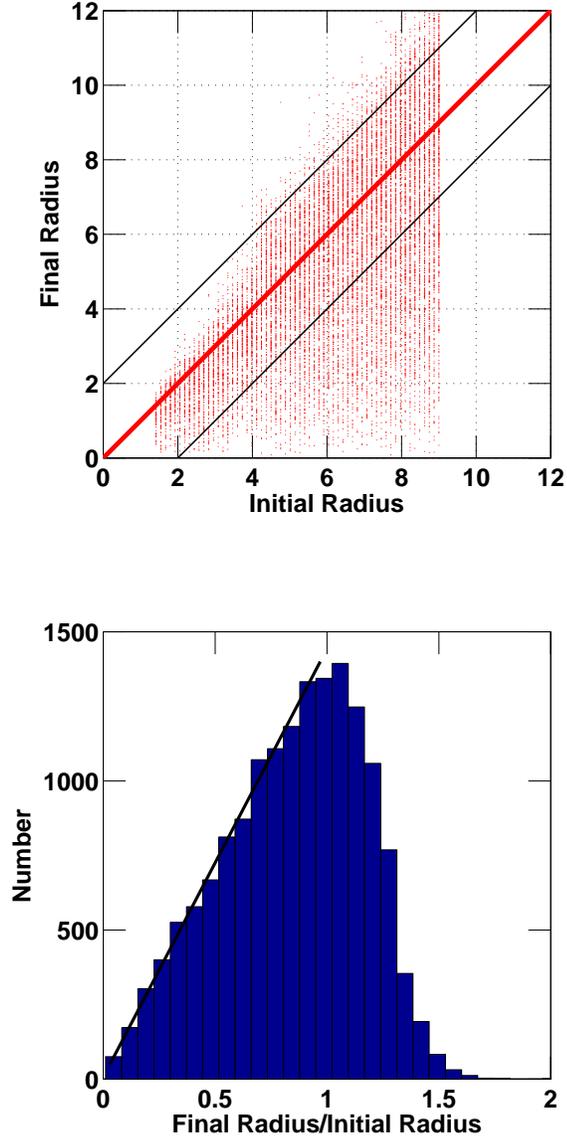}
\caption{(Top) The initial versus final radii of 15,600 particles in a three-dimensional
simulation with a dwarf galaxy potential and star particles scattering off of massive clouds
(Struck \& Elmegreen 2016). The simulation was run for 300 time units (2.94 Gyr). The
thick red line denotes equal initial and final radii, while the thin black lines are
offset by 2 radial units (1 radial unit equals 0.5 kpc). (Bottom) The distribution of
the values of the ratio of final to initial radii in that simulation. A total of 64.5\%
of the particles had a ratio of radii less than 1.0, indicating that an inward scattering
bias naturally occurred in the model. The black line suggests that the
distribution function for this ratio increases approximately linearly as the ratio
increases from 0 to 1.0.
\label{f6}}
\end{figure}

Figure \ref{f6} illustrates this bias in two ways. The top panel shows the radius of all
the stars at the end of the run on the ordinate versus their initial radius on the
abscissa. The simulation time is 300 units, which equals 2.94 Gyr for a dwarf Irregular
potential.  The thick red line of equal initial and final radii separates the stars
scattered to larger radii (above) and smaller radii (below). The thin black lines are
offset by 2 radial units (1 kpc in physical units) from the red line. Most of the stars
scattered outward lie between the red line and the upper black line. Many more stars are
scattered inward, and a substantial fraction of them lie below the lower black line.
Thus, more stars are scattered inward, and many of them are scattered farther inward than
the outward scattering distance of stars scattered outward.

The lower panel of Figure \ref{f6} shows the distribution of the ratio of the final to
initial radii from the top panel. The majority (64.5\%) of the stars have values of this
ratio less than unity, i.e., are scattered inward. Although not shown in these plots,
this inward scattering fraction grows steadily with time and then saturates when the
profile itself saturates to a fixed exponential once the disk gets thick. The black line
in the lower panel is arbitrarily drawn (not a fit) to show the nearly linear form of the
distribution to the left of the peak. The falloff to the right of the peak is steeper and
nonlinear.

The 3D particle scattering model in \cite{struck16} is consistent with the essential
features of the stochastic models described in this paper, namely the exponential
equilibrium distribution of particle positions and the inward scattering bias. A
plausible physical model for this bias is that initially circular orbits become more
eccentric over time as stars scatter off of massive interstellar structures. The stellar
energy hardly changes during such scattering because the structures are much more massive
than each star. Eccentric orbits at fixed energy have less angular momentum than circular
orbits, so the stars gradually lose angular momentum. That loss causes the inward bias.
This situation differs from collective stellar motions induced by torques, such as the
spreading of a disk outside the corotation radius of a bar or spiral, or in the presence
of a companion galaxy. Those collective motions can give stars angular momentum, and they
would presumably happen simultaneously with the more energy-conserving scatterings off of
interstellar structures. Stars are born in nearly circular orbits because the radial
motions of the gas in which they form damp out. Eventually, stellar scattering decreases
because the random motion and disk thickness get large. Then interactions with
interstellar structures become too fast to deflect each star significantly, and too
infrequent with the gas confined to the midplane to deflect each star very often.  This
saturation of the radial profile has the same origin as the usual explanation for the
saturation of the stellar velocity dispersion \citep{lacey84}.  We see this saturation
effect in our 3D simulations as well \citep{struck16}.

\section{Model Tests}
\label{tests}

The stochastic scattering model for exponential profiles has little physical basis at the
moment, and even if it operates in real galaxies, it may be overwhelmed by other
processes and be difficult to see directly. The most direct test might be in other
numerical simulations where small disk perturbations from gas structures or flocculent
spiral arms scatter stars to different epicyclic guiding centers. Our prediction is that
the distribution of the ratio of the guiding center radius after each scattering event to
the radius before that event has a slight asymmetry toward values less than 1. That would
confirm the assumption of an inward bias in scattering probability resulting from small
perturbations.  This is a different distribution function than the ratio of final to
initial radius after many orbits, which will also contain bulk motions and resonant
scattering off global spirals and bars. It is also different from the distribution
function in the outer or inner parts of a disk, which will have a selection bias because
most stars there scattered preferentially outward or inward to get to these positions.

An observational test might be provided by the far-outer regions of spiral galaxies,
beyond the spirals where the disk gets relatively smooth. In two of the three galaxies
observed by \cite{watkins16}, the far-outer emission appeared to come from a disk rather
than a halo of stars.  These outer disk regions are still approximately exponential and
yet there are no stellar features there like flocculent spiral arms that could scatter
stars.  Also in some low surface brightness galaxies there are no obvious spiral arms. A
first test is if the stellar orbits are circular. Any gas that is present should have
more circular orbits than stars because gas orbits cannot cross each other. If the
stellar orbits are eccentric, then the rotation curve for the stars would be more
declining in the outer parts than the gas.  With both corrected for random motions via
the asymmetrical drift correction, they should look the same.  If the uncorrected gas and
stars have the same rotation curves, then the stars are probably in circular orbits. For
example, the low surface brightness galaxy ESO-LV 2340130 in \cite{pizzella08} has a
slightly lower rotation speed for the stars than the gas, suggesting eccentric orbits for
the stars. The effect should generally be small because frequent and weak stellar
scatterings off of numerous, extended gas structures will not make highly eccentric
orbits \citep{struck16}.  Nevertheless, the test would involve mapping the column density
and structure of gas in these spiral-free regions to see if there are significant
perturbations from clouds, shells and holes. In the outer parts of disk galaxies and in
low surface brightness galaxies, gravitational forces are relatively weak, relative
orbital motions are slow, and cloud interaction timescales are long, so the required
surface density irregularities in the gas can be small too and yet stellar scattering
might have an important cumulative effect. If there is no gas in these regions and the
stellar orbits are nearly circular, then something else might be scattering the stars
locally, such as tiny satellite galaxies or dark matter mini-halos.

Exponential profiles in galaxies without gas clouds or other perturbations for stellar
orbits would seem to be remnants of a former time when they had such scattering. Some S0
galaxies might be in this category.

The idealized stochastic model discussed here scatters particles all over the disk.
Migration to outer galaxy disks is observed \citep{yoachim12,roediger12,ruiz16}, but the
model also scatters particles to the inner disk, making an exponential profile from the
outside in if there is initially a gap there. In real galaxies, several processes
operating over a broad range of radii in the inner parts of disks could provide the
reflection needed for the stochastic process. Then migrating stars need not get so close
to the center. However, observations of stars in inner disks that migrated there from
further out would be interesting.

\section{Summary}
\label{sum}

Scattering of particles in 1 dimension produces the usual Gaussian distribution function
when the particle can travel anywhere on the line, but this function changes to an
exponential when there is a bias in the probability toward a reflecting barrier. The
scale length of the exponential is $R_{\rm D}=1/\ln(q/p)$ for inward and outward
probabilities $q$ and $p$ and unit step size.

Scattering in 2 dimensions also produces exponential distribution functions in radius, or
$\exp(-r/r_{\rm D})/r$ type functions which are nearly indistinguishable from
exponentials, when there is an inward bias for each scattering event. Scattering can fill
the whole disk out to very large radii even if the particles are launched from a narrow
range of radii. The exponential reaches an equilibrium shape after a large enough number
of scatterings per particle, with a scale length of $R_{\rm D}=0.5\lambda/(q-p)$ for
scattering mean free path $\lambda$. Double exponentials analogous to galaxy profiles of
Types II and III can be generated by varying the scattering bias over radius.

The scattering bias was determined from a previous three-dimensional numerical simulation
\citep{struck16} in which star particles scattered off of massive diffuse clouds in a
dwarf galaxy potential.  The simulation, which produced an exponential disk as shown in
that paper, had an inward scattering bias shown here by the distribution of ratios of
final to initial radii.

The scattering model may be relevant to the parts of galaxy disks that have been
perturbed by interactions, accretion, ring formation, and other processes that change the
average radial disk structure. Stellar scattering should gradually smooth over these
irregularities, making them tend toward an exponential (or $e^{-r/r_{\rm D}}/r$) shape.
Scattering should also change an initially tapered disk that forms by cosmological
accretion into an exponential or piece-wise exponential. The scattering bias assumed for
the model is plausible because stars born in circular orbits that scatter elastically off
massive clouds and other ISM structures lose angular momentum at nearly constant energy,
which creates an inward bias.

\end{document}